\documentclass[paper]{agujournal2019}
\usepackage{url}
\usepackage{lineno}
\usepackage{color}
\usepackage{amssymb,amsmath}
\usepackage{hyperref} 
\hypersetup{
    colorlinks=true,
    linkcolor=blue,
    filecolor=magenta,      
    urlcolor=cyan,
}

\begin{document}


\title{Searches for Supersymmetry (SUSY) at the Large Hadron Collider}
\authors{Sophie Kadan\affil{1}}
\affiliation{1}{University of Pennsylvania, Philadelphia, PA, USA}
\correspondingauthor{Sophie Kadan}{sokadan@sas.upenn.edu; On behalf of all authors, the corresponding author states that there is no conflict of interest.
}

\begin{abstract}

This paper provides an overview of supersymmetry (SUSY) and the ongoing efforts to detect SUSY particles at the Large Hadron Collider (LHC). SUSY proposes corresponding "sparticles" for each Standard Model particle, with the potential to resolve the hierarchy problem, identify dark matter, and unify forces at high energy scales. We explore the key observables central to SUSY searches at the LHC, such as effective mass (\(M_{\text{eff}}\)), scalar sum of transverse momenta (\(H_T\)), and stransverse mass (\(M_{T2}\))--each of which is instrumental in distinguishing SUSY particles by their mass scales and decay characteristics. The paper also discusses the challenges of differentiating SUSY signals from Standard Model backgrounds, particularly in scenarios without significant missing transverse momentum (\(E_T^{\text{miss}}\)).

A critical aspect of these searches is achieving high signal purity while effectively managing background detection. The LHC employs sophisticated detection techniques to enhance the signal-to-noise ratio--crucial for isolating genuine SUSY events from prevalent Standard Model processes such as multijet productions, gauge bosons with jets, and top quark pairs. Through this examination, the paper aims to illuminate the challenges and potential breakthroughs in SUSY searches, assessing their broader implications for fundamental physics.
\end{abstract}

\section{Introduction}

Supersymmetry (SUSY) is a novel theory that expands upon our current Standard Model of particle physics. While traditional symmetries such as rotational, Lorentz, and gauge symmetries often relate similar types of particles, SUSY is unique in that it establishes a relationship between two fundamentally different classes of particles—bosons and fermions. This theory posits that for every boson, a corresponding fermion exists, and vice versa. This relationship suggests that particles should have supersymmetric counterparts, often referred to as “sparticles,” which share the same mass but differ in spin by a half-integer value [1][2].

One of the key impacts of SUSY is its potential to unify the fundamental forces and particles into a single framework, reconciling Standard Model shortcomings such as the hierarchy problem and the nature of dark matter. Experimental searches for sparticles, such as those conducted at particle accelerators, are therefore crucial in verifying this theory. Many such experiments conducted by the ATLAS group at the Large Hadron Collider (LHC) are designed to detect the unique signatures that sparticles would leave behind if they were produced in particle collisions. For instance, certain supersymmetric models predict the production of sparticles like gluinos [3].

This paper contains six sections. In Section \ref{sec:principles} we delve into the theoretical and experimental underpinnings of supersymmetry, outlining its principles, implications, and potential role in reconciling Standard Model shortcomings. In Section \ref{sec:lhc} we detail the methods and challenges of detecting SUSY, particularly via high-energy collider experiments conducted at the LHC. In section \ref{sec:background} we discuss the process of distinguishing between SUSY signal and SM background, with an emphasis on the detection of key SUSY observables. In Section \ref{sec:interpret} we discuss the interpretation of detector results, comparing the implications of these interpretations in Section 6 after removing assumptions like R-Parity conservation. 

\section{SUSY Principles and Motivations}\label{sec:principles}
\subsection{The Standard Model}
The Standard Model (SM) of particle physics describes fundamental particles and their interactions with remarkable precision. It describes matter as fundamentally composed of various generations of quarks and leptons, each with specific properties such as mass, electric charge, and spin. Likewise, it sufficiently explains particle interactions through the exchange of gauge bosons, which serve as carriers for the strong, weak, and electromagnetic forces. Notably, a recent achievement within this framework has been the discovery of the Higgs boson: by providing a mechanism for particle mass generation, the Higgs boson discovery thus completed the SM [4].

Despite its successes, the SM is nonetheless considered an incomplete representation of the fundamental forces, particularly because it does not address shortcomings such as the hierarchy problem: this problem highlights the unnatural fine-tuning required to keep the Higgs boson mass at its relatively low value despite theoretical predictions that suggest it should be much higher. Likewise, the SM leaves several crucial questions unanswered, such as the origins of dark matter, the matter-antimatter asymmetry in the universe, and the nature of mass. Furthermore, it does not explain the origin of particle quantum numbers, the mechanisms by which neutrinos acquire mass, or offer a unified framework for all interactions that incorporates gravity [3][5].

Supersymmetry (SUSY) stands out as a compelling theoretical extension of the Standard Model because it proposes solutions to many of these unresolved issues. SUSY not only provides a framework that could potentially incorporate gravity with other fundamental interactions but offers an explanation for the origin and composition of dark matter. In addition to addressing the matter-antimatter asymmetry, its inclusion of sparticle interactions offers a compelling resolution to the hierarchy problem. Finally, supersymmetry suggests a way to explore the origins of particle quantum numbers and neutrino masses within a unified theoretical framework, possibly pointing towards a grand unified theory [1][2]. 

\subsection{SUSY Principles}
Supersymmetry (SUSY) is a theoretical model of particle physics that proposes a unique type of symmetry between two fundamental classes of particles: bosons, which are force carriers, and fermions, which constitute matter. In the SUSY framework, each fermion of the Standard Model, such as quarks and leptons, is paired with a corresponding bosonic "sfermion" (e.g.,  (\( \tilde{f}_{L, R} \)) ) with zero spin. These sfermions share the same gauge quantum numbers\footnote{The underlying symmetry of SUSY, wherein its generators commute with those of internal symmetries, is what preserves these crucial gauge quantum numbers. This preservation is key to ensuring that the interactions of the sparticles mirror those of their Standard Model counterparts.}  as their fermionic counterparts, maintaining the balance and symmetry that gauge theories require [5].

SUSY also modifies the role of the Higgs boson by introducing its fermionic partners, termed Higgsinos, which are integrated within the Higgs chiral supermultiplets. Further extending its reach, SUSY associates the gauge fields responsible for mediating the fundamental forces with Majorana fermions known as gauginos.

The pairings that SUSY offers not only preserves the functional symmetry of forces but also ensures an equivalence in the number of bosonic and fermionic degrees of freedom, which is vital for the internal consistency of the model. By doing so, SUSY provides a structured and unified approach to understanding particle interactions, bridging the gap between matter particles and force carriers. Such consistency not only boosts the predictive power of SUSY but also enhances its potential to be empirically verified in high-energy physics experiments[4][5].

\subsection{"Soft" Breaking}
Supersymmetry introduces several distinctive features that set it apart from the Standard Model, notably its ability to accommodate renormalizable interactions that can potentially violate baryon and lepton numbers. This is facilitated by SUSY’s inclusion of scalar sparticles, which could lead to significant phenomena such as proton decay--especially if the masses of these sparticles are below the TeV scale [2].

To manage potential issues such as unwanted rapid decay processes, SUSY is theorized to be a broken symmetry. Known as "soft" breaking, this phenomena is crucial because it mitigates SM fine-tuning problems while preventing the rapid decay phenomena that would otherwise contradict observed physical data. Such an approach suggests that while sparticles theoretically exist, they remain undetected due to their significantly higher-than-expected masses [1].

Soft SUSY is supported by the fact that even light sparticles such as the selectron have not been detected at expected mass scales. These aspects highlight that evidence of supersymmetry may only become apparent at energy scales not yet accessible experimentally [6]. 

\subsection{Motivations for SUSY}

SUSY is appealing because it answers many of the key questions that the Standard Model cannot address alone. Of these, the hierarchy problem continues to pose a significant theoretical challenge to current models of the Higgs boson and its interactions: here, the Higgs boson's mass requires extreme fine-tuning to maintain its observed low value against potential high quantum corrections up to the Planck scale. SUSY offers a compelling solution to this problem by introducing superpartners for each particle: these interactions naturally cancel out large quantum corrections at the TeV scale, which resolves the need for fine-tuning [1][2].

Supersymmetry is also favorable because it provides a potential candidate for dark matter, particularly when R-parity is conserved. R-parity is a crucial quantum number derived from the baryon number (B), lepton number (L), and the spin quantum number (S) of the particles: defined as \( P = (-1)^{3(B-L) + 2S} \), this relation posits that Standard Model particles have \( P = +1 \) while SUSY particles have \( P = -1 \). R-parity Conservation (RPC) is particularly favorable because it implies that SUSY particles can only be produced in pairs. RPC also theorizes a lightest supersymmetric particle (LSP) that is stable, deeming it a suitable dark matter candidate. This stability arises because, with conserved R-parity, the LSP cannot decay into lighter standard model particles alone\footnote{For this exposition, it is assumed that R-parity violating (RPV) couplings are absent or sufficiently small for the LSP's lifetime to exceed the age of the universe. This allows the LSP to still function as a dark matter candidate. Considerations such as these significantly impact the potential detection strategies of SUSY--as well as the theoretical implications of SUSY in providing a comprehensive understanding of dark matter in the universe.} [2][5][6].

Furthermore, supersymmetry inherently incorporates gravitational interactions by predicting the existence of the graviton and its SUSY partner, the gravitino--allowing it to potentially unify gravity with other fundamental forces in a single theoretical framework. Additionally, SUSY predicts that the coupling constants of the strong, weak, and electromagnetic forces converge at a grand unification scale when extrapolated to the early universe, theorizing a unification that is not achieved by the SM alone [4].

These aspects make SUSY a robust theoretical extension of the Standard Model, promising to address fundamental questions about the universe’s structure and the underlying principles governing it.

\section{Searches for Supersymmetry at the LHC}\label{sec:lhc}

The Large Hadron Collider (LHC) plays a pivotal role in the search for SUSY, utilizing advanced experimental techniques to explore beyond the Standard Model. In fact, the detection and analysis strategies it employs are well designed to reveal the even some of the most subtle signatures of SUSY particles [7].

\subsection{The ATLAS Detector}

\begin{figure}
    \centering
    \includegraphics[width=1\linewidth]{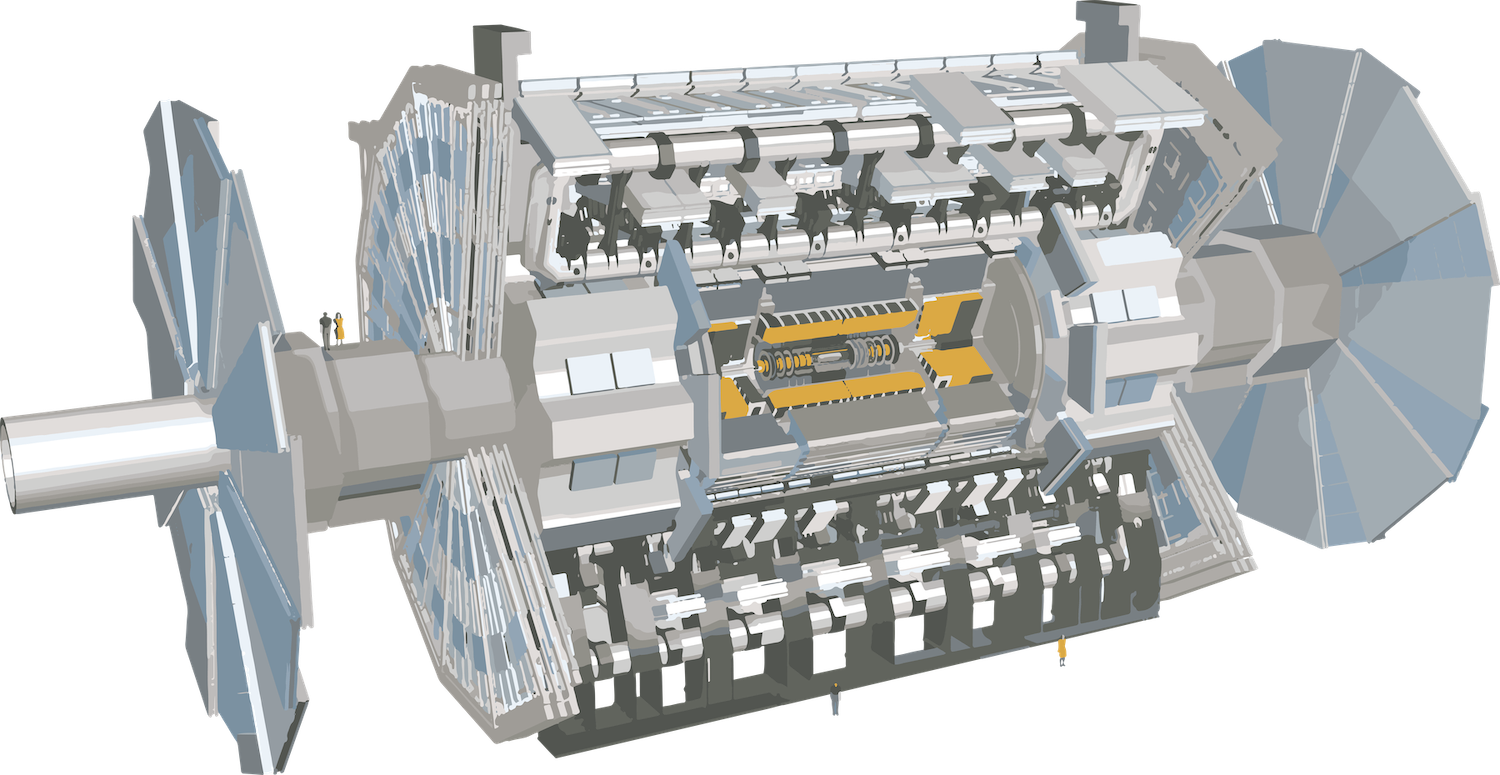}
    \caption{Diagram of the ATLAS Detector, including the components it employs to accurately measure particle interactions [14].}
    \label{fig:det}
\end{figure}

The ATLAS detector at the Large Hadron Collider (LHC) (Figure 1) is designed with a cylindrical geometry that provides near-complete coverage in solid angle, crucial for detecting beyond the Standard Model interactions through proton collisions. This general-purpose detector utilizes a series of systems to identify and measure the properties of particles produced in such high-energy collisions:
\begin{itemize}
    \item Inner Detector: Central to the detector's function, the Inner Detector specializes in reconstructing charged particle tracks in the pseudorapidity region \(|\eta| < 2.5\). Located within a 2 T axial magnetic field, it uses a combination of silicon pixel detectors, silicon microstrip detectors, and a transition radiation tracker (TRT) to measure charged-particle momenta. Likewise, it identifies electron trails via ionization in xenon- or argon-based gases.
    
    \item Calorimeters: Above the Inner Detector, the electromagnetic and hadronic calorimeters span a region up to \(|\eta| < 4.9\), capturing and measuring particle energy. These calorimeters function through layers that absorb and induce particle showers, with interspersed sampling layers quantifying the energy of these showers. Notably, the hadronic calorimeter plays a key role in identifying jets initiated by b-quarks.
    
    \item Muon Spectrometer: For muons, which are typically too massive to leave significant energy in the calorimeters, the muon spectrometer (MS) extends up to \(|\eta| < 2.7\) and tracks muon momenta using its surrounding trigger and tracking chambers in a toroidal magnetic field.
    
    \item ATLAS Trigger System: Comprised of both hardware and software components, the ATLAS Trigger System is essential for reconstructing and analyzing real and simulated data events. It efficiently pinpoints the locations and identities of particles, facilitating the detection of potential new physics phenomena like those predicted by supersymmetry (SUSY) [7].

\end{itemize}

This setup underscores the sophisticated infrastructure required to explore SUSY and other theoretical models at the frontiers of particle physics.

\subsection{Theorized Production of SUSY Particles}

At the LHC, SUSY particles are theorized to be produced primarily through high-energy collisions that potentially generate various types of sparticles. For instance, gluinos and squarks are expected to dominate SUSY production due to their significant roles in SUSY models. Electroweakinos, sleptons, and sneutrinos are also sought after but have smaller production cross-sections. An example where theoretical assumptions play a role in these production dynamics is the conservation of R-parity in many SUSY models, which predicts that sparticles must be produced in pairs. This affects both the production rate and the decay patterns observed.

One of the primary methods of detecting SUSY particles is via the observation of missing transverse momentum (\( \text{E}_{T}^{\text{miss}} \)), which signals the presence of non-interacting particles escaping detection. The measurement of \( \text{E}_{T}^{\text{miss}} \) is crucial, as it often points to the stability and weakly interacting nature of the lightest SUSY particle (LSP). The hallmark of many SUSY searches, a stable LSP typically leaves no detectable signal other than its contribution to missing energy [1][2].

\section{Signal Purity and Background Measurement}\label{sec:background}

The SUSY phenomenology at the LHC is heavily influenced by the underlying properties of the LSP and other sparticles. In fact, the expected detector signatures are determined by how these particles decay and interact with the LHC's detectors. However, distinguishing these SUSY signatures from the background noise created by Standard Model processes poses a far more significant challenge. This is because Standard Model events often mimic SUSY signals, necessitating sophisticated analytical techniques to differentiate between potential discoveries and false positives [8].

\subsection{SM Background}

The difficulty in differentiating SUSY from SM processes is primarily due to the similarities in their decay products. For instance, the SM can result in final states that include multiple jets, leptons, or missing transverse momentum—-all key signatures in SUSY searches. The extent to which a specific SM process interferes with SUSY detection depends on several factors:
\begin{itemize}
    \item Production Cross-Section: This refers to the probability of occurrence for a particular process within the collider. Higher cross-section processes are more common and thus more likely to contribute to background noise in SUSY searches.
    \item Branching Fraction: This metric indicates the likelihood of a process decaying into the specific final state of interest. A higher branching fraction for a decay path similar to a SUSY signal increases the background complexity.
    \item Event Selection: The strategies used to select and analyze events significantly affect how well SUSY signals can be isolated from SM backgrounds. Effective event selection criteria are essential to enhance signal sensitivity and reduce background interference [2][8][9].
\end{itemize}

Presented by the CMS Collaboration, Figure 2 compares current measurements of SM cross-sections with their theoretical predictions. This summary is critical for SUSY search strategies--allowing researchers to understand the landscape of potential background processes and tailor their detection methods accordingly. By analyzing these cross-sections, researchers may prioritize which SM processes might most likely mimic SUSY events and develop more targeted approaches to filter out these backgrounds [1].

\begin{figure}
    \centering
    \includegraphics[width=1\linewidth]{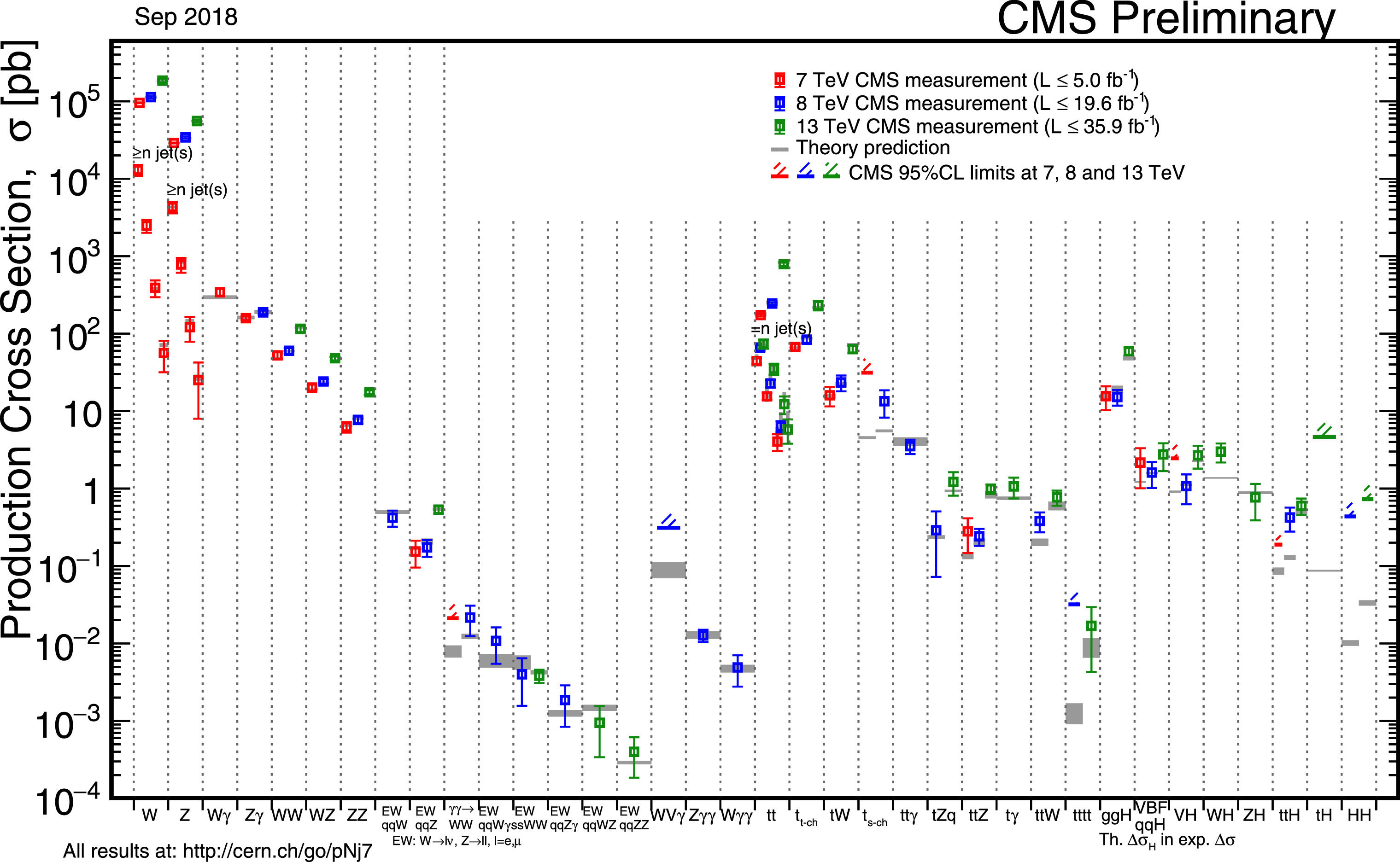}
    \label{fig:2}
    \caption{Current measurements of relevant SM cross-sections and their comparison to theoretical predictions [1].}
    \end{figure}

\subsection{Background Estimation}

To accurately estimate background from Standard Model processes, researchers at the LHC use simulated samples and actual collision data from control samples. The control samples that are analyzed are carefully chosen so that they are representative of specific SM processes but largely free from the signals being searched for. By using these samples, experimentalists may model the backgrounds that mimic SUSY signals more accurately.

For backgrounds that result primarily from strong interactions, collision data is often used due to the significant uncertainties associated with the theoretical cross-sections of these processes. Conversely, backgrounds from electroweak processes are usually estimated using simulated samples because their theoretical predictions are generally more reliable.

Each search develops a tailored method for estimating backgrounds based on the specific event selection strategies employed. This approach is designed to maximize the discovery potential by optimizing the discrimination between the expected signal and the background noise. The ATLAS and CMS collaborations enhance this process by defining "signal regions"--categories within the data characterized by SUSY-like event selection criteria. The classification of such events based on their content and properties is essential for the effective identification of genuine SUSY events amidst the vast data collected at the LHC [1][2][7][8][9].

\subsection{Key SUSY Observables}

SM background is largely dominated by processes like multijet production, gauge bosons associated with jets, and top quark pairs. Such processes typically exhibit only mild missing transverse momentum (\(E_T^{\text{miss}}\)), barring significant mismeasurement of jet signal. Likewise, its decay kinematics are indicative of lower mass particles, with the heaviest particle at about 175 GeV.
To enhance the detection of SUSY particles from such processes, researchers search for several SUSY observables that are correlated to the mass scale of the potential parent particle or the mass differences within the SUSY spectrum:
\begin{itemize}
    \item Effective Mass (\(M_{\text{eff}}\)): Defined as the scalar sum of the transverse momenta (\(p_T\)) of the jets plus \(E_T^{\text{miss}}\). \(M_{\text{eff}}\) is particularly sensitive to events involving heavier particles like gluinos, showing a harder spectrum which helps differentiate these events from SM backgrounds.
    \item Total Transverse Momentum (\(H_T\)): Similar to \(M_{\text{eff}}\), this is the scalar sum of the transverse momenta of all jets, enhancing sensitivity to events with high-energy jets.
    \item Sum of Masses of Large-Radius Jets (\(M_{\text{LRJ}}\)): Focuses on the combined mass of large-radius jets, useful for identifying events where these jets may originate from the decay of heavy particles [9][10].
\end{itemize}

Another critical observable is the stransverse mass (\(M_{T2}\)), a generalization of the transverse mass concept applied to events featuring two invisible particles. \(M_{T2}\) provides a clear endpoint measurement close to the mass of the parent particle--aiding in the distinction from SM processes where such endpoints would not typically appear. Optimal sensitivity is achieved by classifying events based on characteristics such as the presence of b-jets, whether W/Z/h bosons or top quarks are boosted, and the number, charge, and flavor of leptons involved [9][10]. 

\subsection{Complex Final States}

Detecting SUSY is further complicated by additional high cross-section Standard Model processes. QCD multijet production, a common SM process, does not typically include energetic leptons or neutrinos. However, jets can be misidentified as leptons, and missing transverse momentum (\( \text{E}_{T}^{\text{miss}} \)) may arise from mis-measurement of visible particles. Such errors can cause multijet events to falsely resemble SUSY signatures, which often rely on detecting events with leptons and significant \( \text{E}_{T}^{\text{miss}} \).
Electroweak (EWK) production of W and Z bosons and top quark pairs also contribute to the background, particularly when these particles decay into combinations of jets, leptons, and neutrinos, mimicking potential SUSY events. Additionally, rare SM processes like triple gauge boson production or Higgs boson production become relevant due to the LHC’s high energy and large dataset. These processes can produce complex final states with multiple jets, leptons, and \( \text{E}_{T}^{\text{miss}} \), further blurring the lines between SM backgrounds and genuine SUSY signals [11].

\section{Interpreting SUSY Search Results}\label{sec:interpret}

When observed search data aligns with SM background expectations within statistical and systematic uncertainties, the findings are used to set an upper limit on the SUSY production cross-section. This interpretation utilizes likelihood fits under three hypotheses: a background-only model, a model-independent signal plus background, and a model-dependent signal plus background. Each fit incorporates data from both signal and control regions defined in the analysis--reducing systematic uncertainties by constraining expected backgrounds to observed yields [12].

\subsection{Gluino Detection}

Historical data from the Tevatron collider and predictions from various SUSY scenarios suggest that gluinos, a fundamental component in many SUSY models, likely have masses exceeding several hundred GeV. Despite these high mass predictions, gluinos are expected to have sufficiently large cross-sections, making their pair production a classical focus of SUSY searches at high-energy proton-proton colliders. Assuming R-parity is conserved, gluinos are produced in pairs and rapidly decay into final states featuring standard model particles and two neutralinos (denoted as \( \tilde{\chi}^0_1 \)), which typically carry away significant missing transverse momentum (\( \text{E}_{T}^{\text{miss}} \)) due to their high momentum and non-interaction with the detector [1][2][9][10].

\begin{figure}
    \centering
    \includegraphics[width=1\linewidth]{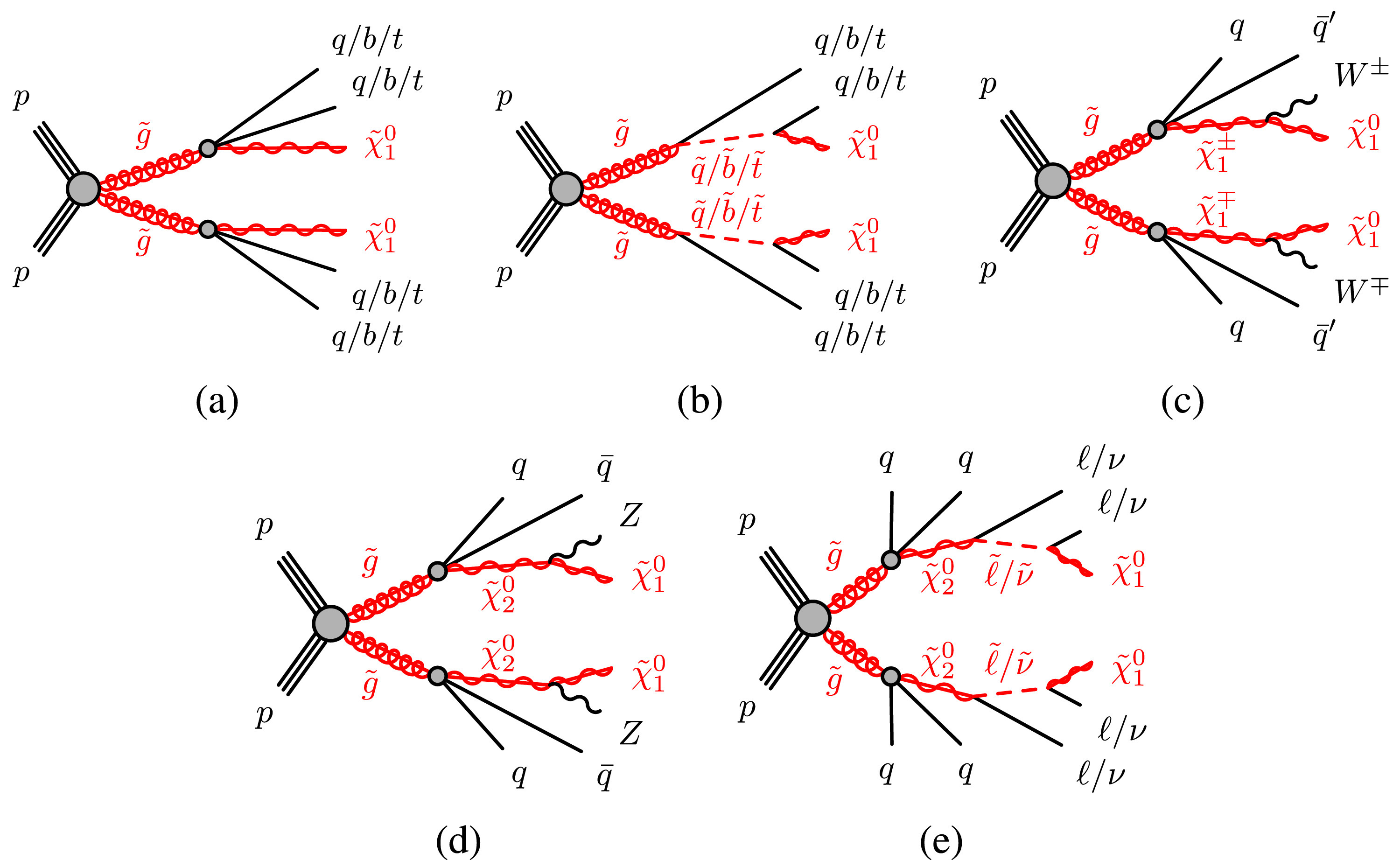}
    \caption{R-parity conserving decays of gluinos under different SUSY spectrum assumptions: (a) When gluinos are lighter than squarks, they decay via virtual squarks, but this is suppressed if squarks are significantly heavier. (b) If gluinos are heavier than squarks, the decay proceeds directly through squarks. (c, d, e) When electroweakinos and sleptons are lighter than gluinos, these provide feasible decay paths [1]}
\end{figure}

The specific decay paths of gluinos depend on the SUSY mass spectrum, which remains largely unknown. However, many decay scenarios are characterized by the production of multiple energetic jets—-resulting from the large mass difference between the gluinos and their decay products—-as well as substantial \( \text{E}_{T}^{\text{miss}} \). This is illustrated in Figure 3, which depicts several R-parity conserving gluino decay diagrams: 
\begin{itemize}
    \item Gluino Mass Smaller Than Squark Mass (Scenario a): In this configuration, the gluino can decay through a process involving virtual squarks. However, if the squark mass is several hundred GeV larger than the gluino mass, this decay mode becomes suppressed. The large mass difference increases the virtuality of the squarks involved, thereby reducing the probability of this decay channel due to the higher energy required to bridge the mass gap.
    \item Gluinos Heavier Than Squarks (Scenario b): This arrangement allows gluinos to decay directly via real squarks, which are physically present and lighter than the gluinos. This mode is typically not suppressed because the real squarks can more readily mediate the decay, leading to a more straightforward gluino decay process compared to the virtual squark scenario.
    \item Electroweakinos and Sleptons Lighter Than Gluinos (Scenarios c, d, e): When electroweakinos (the supersymmetric partners of the electroweak gauge bosons and the Higgs boson) and sleptons (the supersymmetric counterparts of leptons) are lighter than gluinos, these particles provide additional decay channels. Gluinos can decay into these lighter sparticles, facilitating a range of decay processes that contribute to the richness of SUSY phenomenology at colliders [1][2][9][11][12].
\end{itemize}

Such combinations of high-energy jets and missing energy are therefore a hallmark feature in the search for gluino sparticles.

\subsection{Using Jets to Future Search Sensitivity}

As part of ongoing efforts to detect signatures of supersymmetry (SUSY) at the Large Hadron Collider (LHC), the use of jets has become a cornerstone technique in enhancing the sensitivity of these searches. Jets are algorithms that help manage and interpret the complex data from collision events. Large-R jets (R = 1.0), which cover a wide angle (2 radians), effectively consolidate the showers of particles from these decays into fewer, more manageable clusters. This capability is particularly beneficial for reconstructing high-transverse momentum (pT) Higgs bosons, where decay products are tightly collimated. By using large-R jets, the possible combinations of b-quarks needed for accurate Higgs reconstruction are simplified, improving the likelihood of identifying the correct decay chains critical for SUSY detection [4][13][14].

On the other hand, small-R jets (R = 0.4) play a crucial role in more precisely mapping the trajectories of individual b-quarks, especially in lower pT decay scenarios. These jets are adept at capturing more dispersed b-quark products, which are typical in less energetic Higgs decays, allowing for better granularity in analysis. However, they face challenges in high pT conditions where decay products are close together, as they may fail to merge these closely situated particles effectively [20].

Using jets in this way not only enhances the detection of Higgs bosons in SUSY-related events but also significantly reduces the interference from pileup—irrelevant collisions that can obscure the data. By optimizing jet usage, researchers can improve mass resolution and better distinguish SUSY signals from Standard Model backgrounds. This technique is integral to refining the analysis of high-energy collisions at the LHC, aiding in the broader goal of confirming or refuting the existence of supersymmetry through precision measurements and detailed reconstructions of potential SUSY events [13][14].

\section{Alternate Strategies for SUSY Detection}\label{sec:alternate}

\subsection{R-Parity Violating Decays}

Despite extensive searches, ATLAS and CMS have reported no significant deviations from the SM-only hypothesis. Expanding the search, both ATLAS and CMS have also considered models with gluinos that undergo R-parity violating (RPV) decays. In these scenarios, SUSY particles decay entirely into SM particles, resulting in no significant \(E_T^{\text{miss}}\) and thus, kinematics that closely mimic SM backgrounds. This absence of missing energy allows for the full reconstruction of events, leveraging observables like \(M_{\text{LRJ}}\) to examine if large-radius jets could be from the hadronic decay of heavy SUSY particles. These RPV diagrams are shown in Figure 4.

Figure 4 depicts an example decay that, in scenarios where R-parity is violated, gluinos can decay directly into three quarks through interactions governed by the RPV 3-quark couplings, \( \lambda'' \). These decays bypass typical supersymmetric cascade processes and directly produce Standard Model particles, which could enhance detectability at colliders. The strength of the \( \lambda'' \) couplings plays a crucial role in determining the frequency of these decays, providing a unique pathway to potentially observe signs of supersymmetry in environments where R-parity conservation is not assumed [10][11][12].

\begin{figure}
    \centering
    \includegraphics[width=1\linewidth]{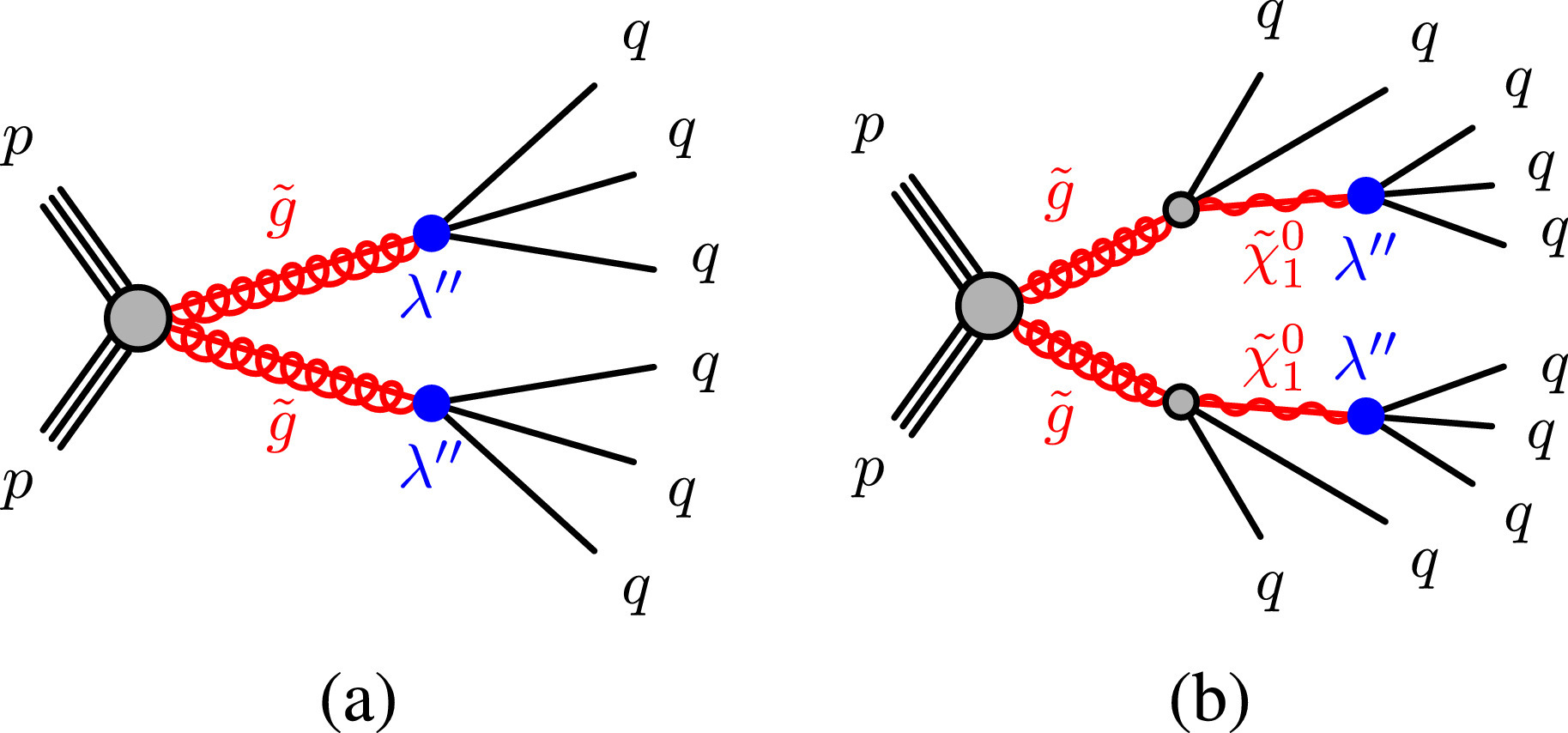}
    \caption{Example R-parity violating decays of gluinos. Here, the strengths depicted are proportional to the RPV 3-quark couplings \( \lambda'' \).}
\end{figure}

\subsection{Targetting R-parity Even Neutral Heavy Higgs Bosons}

Typically, SUSY searches at colliders face significant challenges due to the necessity of pair-producing sparticles when R-Parity is conserved. Not only does the stable LSP often escape detection in these searches, but the need to produce two sparticles simultaneously demands high-energy collisions. This further complicates the direct reconstruction of sparticle masses through traditional resonance techniques--as a particle's properties are often inferred from a peak in an energy distribution that is indicative of its mass [1][2][10].

An alternative to searching for R-parity conserving sparticles focuses on the R-parity even neutral heavy Higgs bosons, specifically the heavy scalar (\(H\)) and the pseudoscalar (\(A\)). This approach offers significant advantages: for one, the particles it searches for can be produced individually through s-channel processes, in which a single particle type is created without the necessity for a counterpart. This mode of production also allows for the direct measurement of their invariant mass—a method akin to how the light scalar Higgs (\(h\)) was detected. Such direct measurement is possible because the heavy Higgs bosons do not typically escape detection, resulting in decay products that can be directly observed and measured [15].

This alternative approach simplifies the detection process at colliders by circumventing the complexities associated with the pair production of sparticles and the issues related to the undetected LSPs. It provides a more straightforward route to probing SUSY manifestations, potentially offering clearer signals and insights into the structure and implications of SUSY [1][15].

\subsection{Co-annihilation and Direct Detection}
Another alternative approach to SUSY detection attempts to address the challenges related to the relic density of dark matter and the observed mass of the Higgs boson. These factors demand precise adjustments within SUSY theories, which typically result in the lightest supersymmetric particle (LSP) being bino-like. However, because bino-like particles do not efficiently annihilate in the early universe, this potentially leads to a relic density that exceeds the bounds established by current observations.

To address this issue, SUSY models often employ the mechanism of co-annihilation. This strategy involves the next-to-lightest supersymmetric particle (NLSP) having a mass very close to that of the LSP, resulting in a compressed spectrum of particle masses. Such proximity allows the LSP and NLSP to effectively co-annihilate, reducing the relic density to within acceptable limits. However, this scenario also complicates detection efforts because the resulting particle decays are typically softer and more difficult to detect, which can obscure signals in experimental data [1][16].

Direct detection experiments are critical in the search for SUSY. Projects like LUX, Panda, and XENON100 have made significant advancements, achieving remarkable sensitivities in measuring spin-independent WIMP-nucleon cross-sections. These experiments aim to detect weakly interacting massive particles (WIMPs) by looking for the tiny recoils that occur when WIMPs collide with nuclei within a detector. Projections indicate that future sensitivity could increase substantially, further enhancing our ability to detect SUSY particles [16][17].

Another promising line of investigation within SUSY searches involves the potential for a higgsino-like neutralino. Unlike a bino-like LSP, a higgsino-like neutralino would likely have higher annihilation rates, which could more naturally align with the observed relic density of dark matter. However, this scenario may require the introduction of additional types of dark matter to fully explain the dark matter composition observed in the universe [2][16][17][18].

These varied experimental approaches highlight the complex, multi-dimensional strategies required to explore the landscape of supersymmetry. Each technique not only contributes to the overarching search for SUSY but also deepens our understanding of potential new physics beyond the Standard Model.

\subsection{Additional Efforts}
If SUSY exists, it faces two main scenarios: either the supersymmetric particles are too heavy to be produced by current accelerators, as some theorists suggest, or they are produced at the LHC but evade detection due to their subtle signatures.

To address the latter case, researchers are now exploring new models that produce exotic or experimentally challenging signatures. These signatures may have been overlooked in previous searches because traditional methods assumed that new particles would decay immediately at the interaction point. However, some unconventional SUSY theories propose the existence of long-lived sparticles that can travel significant distances—from a few micrometers to hundreds of kilometers—before decaying. Both ATLAS and CMS experiments are actively searching for such long-lived particles [19].

Moreover, efforts are underway to detect sparticles that decay into low-energy Standard Model particles. These searches are particularly difficult because they cannot rely on standard detection techniques. Given the abundance of low-energy particles in detector backgrounds, researchers must therefore develop innovative methods to distinguish potential SUSY interactions from ordinary particle noise--making this area of research both challenging and critical for advancing our understanding of particle physics [1][19][20].

\section{Conclusion}
While some in the particle physics community have shifted their focus away from supersymmetry, many experimentalists today remain optimistic about the capabilities of their detectors in searching for SUSY. Further developments are also opening new avenues for analysis and experimentation, preparing researchers for the next phases of data collection. Looking ahead, a High-Luminosity LHC is planned for the late 2020s--promising further enhancement of our exploratory capacity [1][2][13]. 

In the broader scope of future collider projects, many researchers now posit that the motivation for constructing new, higher-energy colliders is predominately focused on Higgs boson studies: these perspectives underline a strategic pivot in high-energy physics towards refining our understanding of the Higgs boson, while keeping the door open for unexpected discoveries in areas like supersymmetry.

\bibliography{full1}

@article{1,
  author = {Canepa, A.},
  title = {Searches for supersymmetry at the Large Hadron Collider},
  journal = {Reviews in Physics},
  volume = {4},
  pages = {100033},
  year = {2019},
  doi = {10.1016/j.revip.2019.100033},
  url = {https://doi.org/10.1016/j.revip.2019.100033}
}

@book{4,
  author = {Ryder, L. H.},
  title = {Encyclopedia of mathematical physics},
  publisher = {VDOC},
  year = {2006}
}

\begin{enumerate}
    \item Canepa, A. (2019). Searches for supersymmetry at the Large Hadron Collider. \textit{Reviews in Physics, 4}, 100033. \url{https://doi.org/10.1016/j.revip.2019.100033}
    \item Tata, X. (1997). What is supersymmetry and how do we find it? \textit{arXiv preprint arXiv:hep-ph/9706307}.
    \item Kane, G. (2017). \textit{Modern elementary particle physics} (2nd ed.). Cambridge University Press.
    \item Ryder, L. H. (2006). \textit{Encyclopedia of mathematical physics}. VDOC.
    \item Schaefer, L. (2019). A search for wino pair production with B-L R-parity violating chargino decay to a trilepton resonance with the ATLAS experiment.
    \item Resseguie, E. (2019). Electroweak physics at the Large Hadron Collider with the ATLAS detector: Standard model measurement, supersymmetry searches, excesses, and upgrade electronics.
    \item Dumitru, S., Herwig, C., \& Ovrut, B. A. (2019). R-parity violating decays of bino neutralino LSPs at the LHC. \textit{Journal of High Energy Physics, 2019}, 1-49. \url{https://doi.org/10.1007/JHEP07(2019)044}
    \item Aad, G., et al. (2021). Search for trilepton resonances from chargino and neutralino pair production in s= 13 TeV pp collisions with the ATLAS detector. \textit{Physical Review D, 104}, 112003. \url{https://doi.org/10.1103/PhysRevD.104.112003}
    \item ATLAS Collaboration. (2012). Observation of a new particle in the search for the Standard Model Higgs boson with the ATLAS detector at the LHC. \textit{Physics Letters B, 716}, 1-29. \url{https://doi.org/10.1016/j.physletb.2012.08.020}
    \item Martin, B., \& Shaw, G. (2001). \textit{Encyclopedia of physical science and technology} (3rd ed.). Academic Press Inc.
    \item Aaboud, M., et al. (2017). Search for supersymmetry in final states with two same-sign or three leptons and jets using 36 fb$^{-1}$ of $\sqrt{s} = 13$ TeV pp collision data with the ATLAS detector. \textit{Journal of High Energy Physics, 2017(9)}, 1-45. \url{https://doi.org/10.1007/JHEP09(2017)084}
    \item Dercks, D., Dreiner, H., Krauss, M. E., et al. (2017). R-parity violation at the LHC. \textit{European Physical Journal C, 77}, 856. \url{https://doi.org/10.1140/epjc/s10052-017-5414-4}
    \item Csáki, C., Grossman, Y., \& Heidenreich, B. (2012). Minimal flavor violation supersymmetry: A natural theory for R-parity violation. \textit{Physical Review D, 85}, 095009. \url{https://doi.org/10.1103/PhysRevD.85.095009}
    \item ATLAS Collaboration. (n.d.). Muon spectrometer. Accessed April 4, 2024. \url{https://atlas.cern/discover/detector/muon-spectrometer}
    \item Ashry, M., Khalil, S., \& Moretti, S. (2023). Searching for a heavy neutral CP-even Higgs boson in the BLSSM at the LHC Run 3 and HL-LHC. \textit{arXiv preprint arXiv:2305.11712}.
    \item Chakraborti, M., Heinemeyer, S., Saha, I., \& Schappacher, C. (2022). (g-2)\_$_\mu$ and SUSY dark matter: direct detection and collider search complementarity. \textit{The European Physical Journal C, 82}. \url{https://doi.org/10.1140/epjc/s10052-022-10414-w}
    \item Beskidt, C., de Boer, W., Hanisch, T., Ziebarth, E., Zhukov, V., \& Kazakov, D. I. (2011). Constraints on supersymmetry from relic density compared with future Higgs searches at the LHC. \textit{Physics Letters B, 695}(1-4), 143-148. \url{https://doi.org/10.1016/j.physletb.2010.10.048} \\
    \item Schaefer, L. (2019). A search for wino pair production with B-L R-parity violating chargino decay to a trilepton resonance with the ATLAS experiment.
    \item Salam, G. P., Wang, L. T., \& Zanderighi, G. (2022). The Higgs boson turns ten. \textit{Nature, 601}, 41–47. \url{https://doi.org/10.1038/s41586-021-04305-9}
    \item Hershberger, S. (2021). The status of supersymmetry. \textit{Symmetry Magazine}. \url{https://www.symmetrymagazine.org/article/the-status-of-supersymmetry}
\end{enumerate}

\end{document}